\documentclass[fleqn,usenatbib]{mnras}
\usepackage{newtxtext,newtxmath}
\usepackage[T1]{fontenc}
\usepackage{subcaption}
\DeclareRobustCommand{\VAN}[3]{#2}
\let\VANthebibliography\thebibliography
\def\thebibliography{\DeclareRobustCommand{\VAN}[3]{##3}\VANthebibliography}

\usepackage{graphicx}	% Including figure files
\usepackage{amsmath}	% Advanced maths commands
\usepackage{xspace}
\newcommand{\x}{x}
\newcommand{\lya}{Ly-$\alpha$\xspace}

\newcommand{\hinvMpc}{\,h^{-1}\, {\rm Mpc}\,}
\newcommand{\hMpcinv}{\,h\, {\rm Mpc}^{-1}\,}

\def\ltsima{$\; \buildrel < \over \sim \;$}
\def\gtsima{$\; \buildrel > \over \sim \;$}
\def\simlt{\lower.5ex\hbox{\ltsima}}
\def\simgt{\lower.5ex\hbox{\gtsima}}

\title[\texttt{MAPLE}: 3D Ly-$\alpha$ Power Spectrum and Covariance Estimation]{Maximum A Posteriori Ly-$\alpha$ Estimator (\texttt{MAPLE}): Band-power and covariance estimation of the 3D Ly-$\alpha$ forest power spectrum}

\author[Horowitz, de Belsunce \& Luki\'c]{
Benjamin Horowitz,$^{1,2}$\thanks{E-mail: \href{mailto:bahorowitz@lbl.gov}{bahorowitz@lbl.gov}}
Roger de Belsunce,$^{2,3}$
Zarija Luki\'c$^{2}$
\\
$^{1}$Kavli IPMU (WPI), UTIAS, The University of Tokyo, Kashiwa, Chiba 277-8583, Japan\\
$^{2}$Lawrence Berkeley National Laboratory, One Cyclotron Road, Berkeley CA 94720, USA\\
$^{3}$Berkeley Center for Cosmological Physics, Department of Physics, University of California, Berkeley, CA 94720, USA
}

\date{Accepted XXX. Received YYY; in original form ZZZ}

\pubyear{2024}

\begin{document}
\label{firstpage}
\pagerange{\pageref{firstpage}--\pageref{lastpage}}
\maketitle

\begin{abstract}
We present a novel maximum a posteriori estimator to jointly estimate band-powers and the covariance of the three-dimensional power spectrum (P3D) of \lya forest flux fluctuations, called \texttt{MAPLE}. Our Wiener-filter based algorithm reconstructs a window-deconvolved P3D in the presence of complex survey geometries typical for \lya surveys that are sparsely sampled transverse to and densely sampled along the line-of-sight. We demonstrate our method on idealized Gaussian random fields with two selection functions: (i) a sparse sampling of 30 background sources per square degree designed to emulate the currently observing the Dark Energy Spectroscopic Instrument (DESI); (ii) a dense sampling of 900 background sources per square degree emulating the upcoming Prime Focus Spectrograph Galaxy Evolution Survey. Our proof-of-principle shows promise, especially since the algorithm can be extended to marginalize jointly over nuisance parameters and contaminants, i.e.~offsets introduced by continuum fitting. Our code is implemented in \texttt{JAX} and is publicly available on GitHub.
\end{abstract}


%%%%%%%%%%%%%%%%%%%%%%%%%%%%%%%%%%%%%%%%%%%%%%%%%%

%%%%%%%%%%%%%%%%% BODY OF PAPER %%%%%%%%%%%%%%%%%%

\section{Introduction}
\label{sec:intro} 
Clustering statistics of large-scale structure tracers carry a wealth of information about dark energy and dark matter. For cosmological inference, the observed data vector is usually compressed to a summary statistic specific to the observed tracer. % to make it computationally manageable, average over stochastic fluctuations and reduce noise. 
For the \lya forest flux fluctuations, a series of absorption features in quasar and galaxy spectra due to intervening neutral hydrogen in the highly ionized intergalactic medium (IGM) along the line of sight \citep{scheuer1965sensitive,gunn1965density}, two main approaches have been used for cosmological data analysis: 

First, the dense sampling along the line-of-sight motivated measuring the one-dimensional, or line-of-sight, \lya power spectrum \citep{1999croft,2002croft,2004MNRAS.347..355K,2005ApJ...635..761M,2013A&A...559A..85P,Chabanier:2019,Pedersen:2020,Karacayli2022,2023MNRAS.526.5118R}. It is a unique probe which puts tight constraints on dark matter models from the small scale clustering 
\citep{Viel:2013,Villasenor:2023,Irsic:2024}, non-minimal cosmological models \citep{2018JCAP...09..011G,2023PhRvL.131t1001G}, exotic physics models (i.e. \citet{Viel:2013,2017PhRvD..96b3522I}) and, on larger scales, constrains the $\Lambda$CDM model \citep{2013A&A...559A..85P,2017JCAP...06..047Y}. 

Second, the sparse sampling of the \lya forest transverse to the line-of-sight leads to adoption of the two-point correlation function (2PCF) to measure the baryonic acoustic oscillation (BAO) feature to constrain the expansion history of our Universe \citep{1970Ap&SS...7....3S,1970ApJ...162..815P,2003ApJ...598..720S, 2011JCAP...09..001S,2013JCAP...04..026S,McDonald:2007,Slosar2013,Busca:2013,dMdB:2020} and the broadband shape of 3D correlations \citep{Slosar2013, Cuceu:2021,Cuceu:2023,Gordon:2023}. The \lya forest is a treasure trove of cosmological information on the large-scale distribution of matter in the Universe and probes the underlying matter density at Mpc scales and below at high redshifts ($2\simlt z \simlt 5$). 

A pioneering study on the P3D from realistic simulations \citep{Font-Ribera:2018} and recent measurements of the 3D \lya power spectrum \citep{karim2023measurement,Belsunce:2024} together with advancements in the theoretical modeling of the \lya forest power spectrum \citep[see e.g.,][and references therein]{Ivanov:2023}, have sparked interest in the three-dimensional power-spectrum containing information across lines of sight -- a case in point for our work. Whilst both summary statistics, the 2PCF and P3D, contain, in principle, the same information, both have their advantages \citep{Font-Ribera:2018}: The power spectrum isolates slowly varying features at large scales while having (potentially) less correlated errors. The 2PCF can be calculated using simple pixel-product algorithms and can isolate effects at specific separations, like the BAO peak. Whilst the 2PCF can easily deal with the window matrix, grid-based FFT estimators for the P3D are strongly affected by the sparse sampling of the \lya forest \citep{Belsunce:2024}.

In this paper, we present the Maximum A Posteriori Ly-$\alpha$ Estimator, \texttt{MAPLE}, which can efficiently estimate the 3D \lya power spectrum in band-powers with a corresponding covariance matrix from idealized Gaussian random fields (GRF) with non-trivial survey geometries and varying number densities, typical for \lya forest surveys. To test the accuracy of the band-power reconstruction, we emulate two background source density configurations\footnote{Note that we focus on the effect of the window matrix and ignore contaminants such as the non-Gaussian resolution matrix, continuum fitting, metal contamination, damped \lya absorbers and broad absorption lines.}: (i) the Dark Energy Spectroscopic Instrument (DESI, \citet{DESI:2016}) with $\sim 30$ background sources per $\mathrm{deg}^2$; and (ii) the Prime Focus Spectrograph (PFS, \cite{2022PFSGE}) with $\sim 900$ background sources per $\mathrm{deg}^2$. 

Our estimator expands on earlier work for power spectrum and covariance estimation in more general contexts \citep{2017Seljak,2019Horowitz}, which we extend to the three-dimensional case for the present work. We utilize a modern implementation of a response-type formalism, known as the Marginal Unbiased Score Estimator (MUSE) \citep{2022Muse,2022MUSEII}, to estimate the band-powers and associated covariance matrices.% (i.e. the survey window function). 
We implement our code in \texttt{JAX}, which enables modern implicit differentiable models and optimization methods for rapid estimation even in high dimensional spaces, and supports graphical processor unit (GPU) architecture. We make our implementation publicly available.\footnote{\url{https://github.com/bhorowitz/MAPLE}}

The remainder of this paper is organized as follows: We present the 3D power spectrum estimator in Sec.~\ref{sec:methods}, before discussing the theoretical modeling of the \lya flux power spectrum in Sec.~\ref{subsec:fidpower}. In Sec.~\ref{sec:simulations} we describe our idealized Gaussian random fields and compare the reconstructed band-powers to the theory input power spectrum in Sec.~\ref{sec:results}. In  Sec.~\ref{sec:conclusion} presents our conclusions.

\section{Methods} \label{sec:methods}
We present a novel approach to compress the \lya forest flux fluctuations to the 3D power spectrum using a minimization-based approach. We briefly outline the methodology and focus on \lya-specific implementation details and refer the reader to \citet{2017Seljak,2019Horowitz} for a fuller presentation. 

\subsection{Marginal Unbiased Score Estimation}
\label{subsec:muse}
We want to find the conditional probability distribution of our band-powers, $\theta$, given some observed data, $\x$, which we express as $\mathcal{P}(\x\,|\,\theta)$. However, the data itself depends on $N$ latent parameters, $z$. From those, in turn, some depend on the phases, $\theta$, and others are nuisance parameters and, thus, do not depend on $\theta$. This hierarchical Bayesian problem can be expressed as a marginalization over the latent space parameters,  
\begin{equation}
        \label{eq:hi}
    \mathcal{P}(\x\,|\,\theta) 
    =\int \!{\rm d}^N\!z \; \mathcal{P}(\x,z\,|\,\theta) = \int \!{\rm d}^N\!z \; \mathcal{P}
    (\x\,|\,z,\theta)\mathcal{P}(z\,|\,\theta)\ ,
\end{equation}
where $\mathcal{P}(\x,z\,|\,\theta)$ is the joint likelihood of the data and the latent space parameters, given a fixed $\theta$. In generic cases, this integral is analytically intractable and approximating it relies on Monte Carlo based sampling. More recently, methods have been developed to approximate this integral expression, including the Marginal Unbiased Score Estimation (MUSE) which is the focus of this work \citep{2017Seljak,2019Horowitz,2022Muse}. The core idea of MUSE is to reformulate this integral as an optimization problem for $\mathcal{P}(\x,z\,|\,\theta)$ for fixed $\theta$ and $\x$, and iteratively solve for $\theta$. 

More concretely, at a fixed value of $\theta$ we want to find the maximum a posteriori value (MAP) for $z$ defined by
\begin{align}
    \label{eq:zmap}
    \hat z(\theta,x) \equiv \underset{z}{\rm argmax} \; \log \mathcal{P}(\x,z\,|\,\theta).
\end{align}
It can be shown that the derivatives of the likelihood function at the MAP estimate with respect to the parameters of interest, known as the `score' are optimal statistics in the sense that they preserve all information content of the data \citep{2018MNRAS.476L..60A}. This score function can be expressed as
\begin{align}
    \label{eq:smap}
    s_i^{\rm MAP}(\theta,\x) \equiv \frac{d}{d\theta_i}\log \mathcal{P}(\x,\hat z(\theta,x)\,|\,\theta).
\end{align}
We define the MUSE estimate, $\bar\theta$ as the value of $\theta$ which satisfied the following equation
\begin{align}
    s_i^{\rm MAP}(\theta,\x) - \Big \langle s_i^{\rm MAP}(\theta,\x) \Big \rangle_{x\sim\mathcal{P}(x\,|\,\theta)} = 0,
    \label{eq:musesoln}
\end{align}
where the second term is an ensemble average over realizations of the data drawn from $\mathcal{P}(x\,|\,\theta)$. This ensemble average can be viewed as a generalization of a noise bias which includes all possible sources of biasing from the forward model (i.e. survey geometry).

We can define the covariance matrix associated with this estimate as $\Sigma\,{=}\,H^{-1} J H^{\dagger}$, where $J$ and $H$ are defined as

\begin{align}
\begin{split}
        J_{ij} &= \Big\langle s_i^{\rm MAP}(\bar\theta,\x) \, s_j^{\rm MAP}(\bar\theta,\x) \Big\rangle_{\x\sim\mathcal{P}(\x\,|\,\bar\theta)}  \\ 
        &-\Big\langle s_i^{\rm MAP}(\bar\theta,\x) \Big\rangle \Big \langle s_j^{\rm MAP}(\bar\theta,\x) \Big\rangle_{\x\sim\mathcal{P}(\x\,|\,\bar\theta)}, \label{eq:J}
        \end{split}
\end{align}
\begin{equation}
    H_{ij} = \left. \frac{d}{d\theta_j} \left[ \Big\langle s_i^{\rm MAP}(\bar\theta,\x) \Big\rangle_{\x\sim\mathcal{P}(\x\,|\,\theta)} \right] \right|_{\theta=\bar\theta}. \label{eq:H}
\end{equation}
For a Gaussian likelihood, it can be shown that this definition provides directly the marginal maximum
likelihood estimate and the covariance, $\Sigma$, becomes the inverse Fisher information matrix. Moreover, for a non-Gaussian likelihood, it has been shown that this definition provides an unbiased estimate, although the covariance may not be optimal  \citep{2022Muse}. Note that we can also use this approach to generalize Fisher forecasts, wherein we do not perform the optimization discussed in Eq.~\eqref{eq:musesoln} but instead assume a fixed value of the target quantity $x$. 

We base our analysis code on the \texttt{JAX} implementation of MUSE\footnote{\url{https://github.com/marius311/muse_inference}} discussed in \citet{2022Muse} using the implicit differentiation approach \citep{2022MUSEII} for the Hessian/Jacobian calculations in Eq.~\eqref{eq:J} and \ref{eq:H}.

\subsection{Cosmological Wiener Filtered Tomographic Reconstruction}
\label{subsec:wf}

A key step in the MUSE procedure described above is the inference of the latent space parameters, $z$ for given observed data $x$. This requires a generative response model which may depend on fixed parameters, $\theta$, i.e. $R_\theta$. The generic optimal solution for this problem is the Wiener filter which, in the linear case, can be expressed in terms of the noise covariance $N$ and signal covariance $S$, as 
\begin{equation}
\centering
    \hat{s} = SR_\theta^\dagger(S+N)^{-1}x.
\end{equation}
In the case of the \lya forest (as well as many other large-scale structure probes), the signal covariance is assumed to be that of a fiducial cosmological model and the noise covariance is dominated by detector noise. The former is diagonal in Fourier space, while the latter is near diagonal in real space. Therefore, the resulting matrix inversion of $(S+N)$ is computationally demanding and often is recast as an optimization problem \citep{2017Seljak,2019Horowitz} with associated $\chi^2$ of 
\begin{equation}
\label{eq:opt}
    \chi^2(s_i) = s_i^\dagger S^{-1} s_i + (R_\theta s_i - d)^\dagger N^{-1} (R_\theta s_i -d),
\end{equation}
where $s_i$ is the guess at the $i$th optimization step. 

This approach to Wiener Filtering is in contrast to the smooth priors used in many \lya tomography works \citep{2008arXiv0801.4335C,2018CLAMATO1,2020LATIS,2022WEAVEQSO,2022CLAMATODR2}. These methods are likely more optimal for cosmic web studies, where non-linear non-Gaussian structures are of interest, but smoothness priors are difficult to consistently interpret for cosmological analysis within a Bayesian framework.

We implement Eq.~\eqref{eq:opt} as an optimization problem utilizing particle-mesh routines developed in \citet{2022pmwd}. In our case, the latent space parameters $z$ are the phases of the three-dimensional field, although we note it could be expanded to include nuisance parameters (e.g. continuum fitting parameters) which we leave to future work. The response function, $R_\theta$, can be broken down into the following steps;
\begin{enumerate}
    \item The phases, $z$, are convolved with a transfer function which is determined by the band-powers, $\theta$.
    \item The resulting field is inverse Fourier transformed to real space.
    \item The values of the real space field are read out at the locations of mock observed data using a linear weighting scheme (i.e.  Cloud-in-Cell readout \citep{laux1995particle}) of nearby cells.
\end{enumerate}

\section{Theory Power Spectrum}
\label{subsec:fidpower}
To model the theory power spectrum, we employ the Kaiser formula \citep{Kaiser1987} with a non-linear correction term obtained from hydrodynamical simulations \citep[see ][]{2003ApJ...585...34M, Arinyo:2015}.  In order to perform our analysis we need to choose a basis/binning for our band-powers, $\theta$.  We emphasize that an advantage of our approach is that we  apply the power spectrum as a transfer function on the full density field, meaning that this approach is agnostic to the chosen power spectrum basis.  We define our power spectra in binned $k_\perp$, $k_\parallel$ space with a constant redshift across the bin.

For our proof-of-concept, we focus on the usual redshift-space formula derived from the linear theory of gravitational collapse \citep{Kaiser1987} which includes effects obtained from fits to hydrodynamical simulations \citep{2003ApJ...585...34M};
\begin{equation}
    P_{\rm F}(k,\mu) = b^2 (1+ \beta \mu^2)^2 P_{\rm L}(k)D(k,\mu)\ ,
    \label{eq:p3d}
\end{equation}
where $P_{\rm F}(k,\mu)$ is the flux power spectrum as a function of wavevector $k$ and angle $\mu \equiv k_{\parallel}/k$, $b$ is the redshift-dependent linear bias parameter, $\beta$ the redshift-space distortion (RSD) parameter, $P_{\rm L}(k)$ the linear power spectrum\footnote{We compute it using the Boltzmann solver \texttt{CAMB} (\url{https://camb.info/}).} and $D(k,\mu)$ a non-linear correction term:
\begin{equation}
\label{eq:dkmu}
    D(k,\mu) = \exp \left( \left[\frac{k}{k_{\rm NL}}\right]^{\alpha_{\rm NL}}-\left[\frac{k}{k_{\rm P}}\right]^{\alpha_{\rm P}}-\left[\frac{k_{\parallel}}{k_{\rm V}(k)}\right]^{\alpha_{\rm V}} \right)\ ,
\end{equation}
with 
\begin{equation}
    k_{\rm V}(k) = k_{\rm V0}\left(1+\frac{k}{k'_V}\right)^{\alpha'_V}\ .
\end{equation}

This formula accounts for effects from non-linear evolution (NL), pressure smoothing (P), and velocity effects (V). We choose the values obtained from hydrodynamical simulations \citet{2003ApJ...585...34M} for $\{b^2, \beta, k_{NL}, \alpha_{NL},k_P,\alpha_P,k_{V0},\alpha_V,k'_V, \alpha'_V \} = \{0.0173,1.58, 6.40, 0.569, 15.3, 2.01, 1.220, 1.50, 0.923, 0.451\}$. We set the cosmological parameters to be $\{\Omega_b,\Omega_m, h, A_s \times 10^9,n_s\} = \{0.02214, 0.1414,  0.719, 2.2, 0.961\}$.

\section{Gaussian random field simulations} \label{sec:simulations}

\begin{figure}
    \centering
    \includegraphics[width=0.95\linewidth]{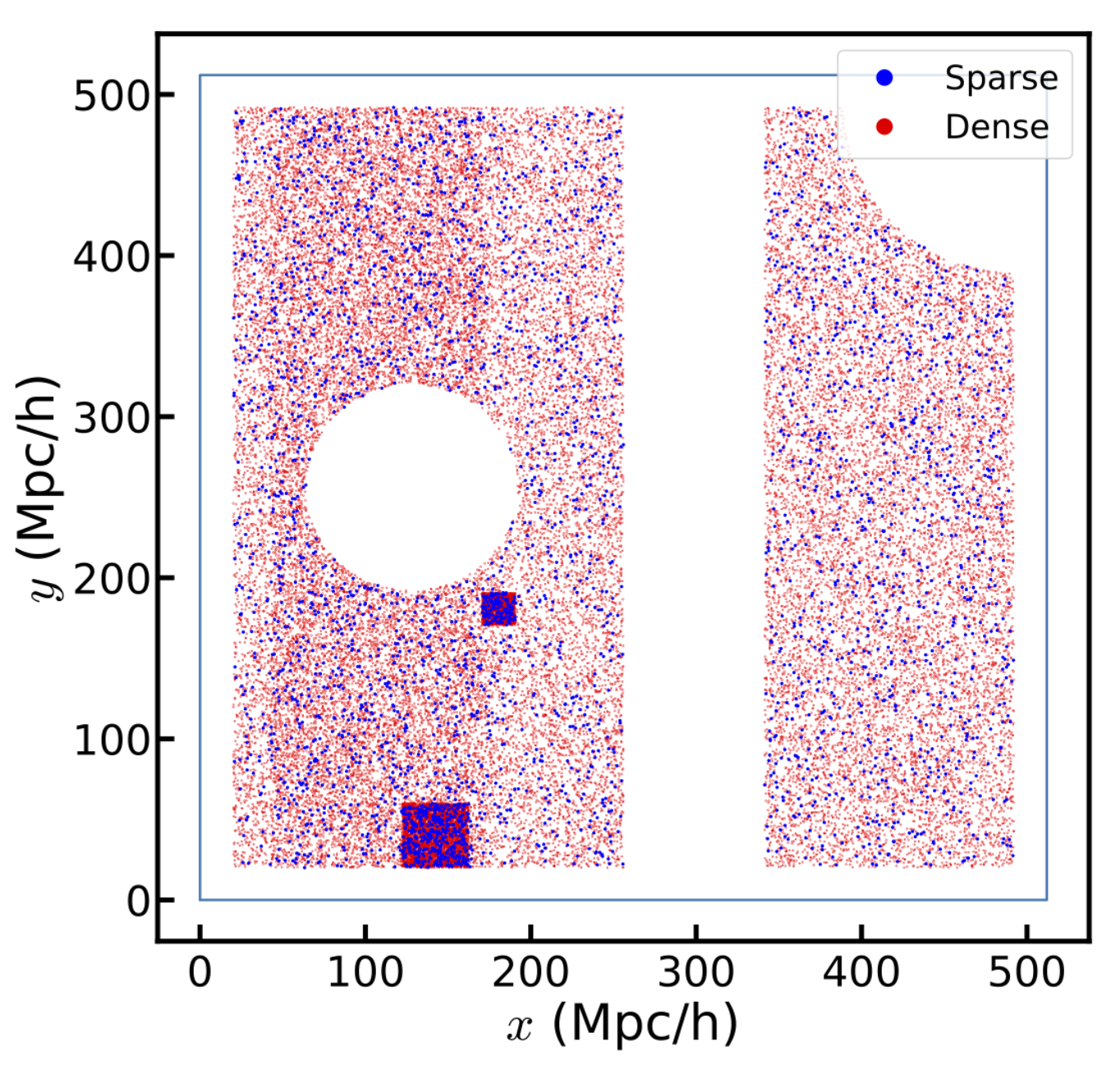}
    \vspace{-0.1in}
    \caption{Mask used for the Gaussian random field for the dense (red) and sparse (blue) configurations, including irregularly masked areas and deep fields. We reconstruct an additional $10\hinvMpc$ on each edge to minimize boundary edge effects.}
    \label{fig:angular_selection}
    \vspace{-0.1in}
\end{figure}

We test \texttt{MAPLE} on an idealized Gaussian realization of our fiducial power spectrum on a $512 \hinvMpc$ box, with resolution of $128^3$. This allows us to reconstruct up to $k_{\rm max} = 0.81 \hMpcinv$.\footnote{The resolution for our MAP reconstruction is implicitly set by $k_\textrm{max}$.} We use 11 bins, logarithmically spaced from $k_{\rm min}=0.001 \hMpcinv$ to $k_{\rm max}$ for both $k_\perp$ and $k_\parallel$. The choice of bin number to use depends on the range of interest, noise properties of the observations, and the volume/resolution of the reconstructed volume. We choose this binning to allow identification of the large scale turnover and the small scale suppression of power. We perform our analysis at a constant fiducial redshift of $z=2.0$.\footnote{In practice, each redshift bin would require a separate MUSE optimization. One could also restructure the procedure described in Sec \ref{subsec:wf} to treat sub-volumes relating to each redshift bin as having different band-power values (i.e. including a varying third dimension to the transfer function).}

\begin{figure*}
    \centering
    \includegraphics[width=1\linewidth]{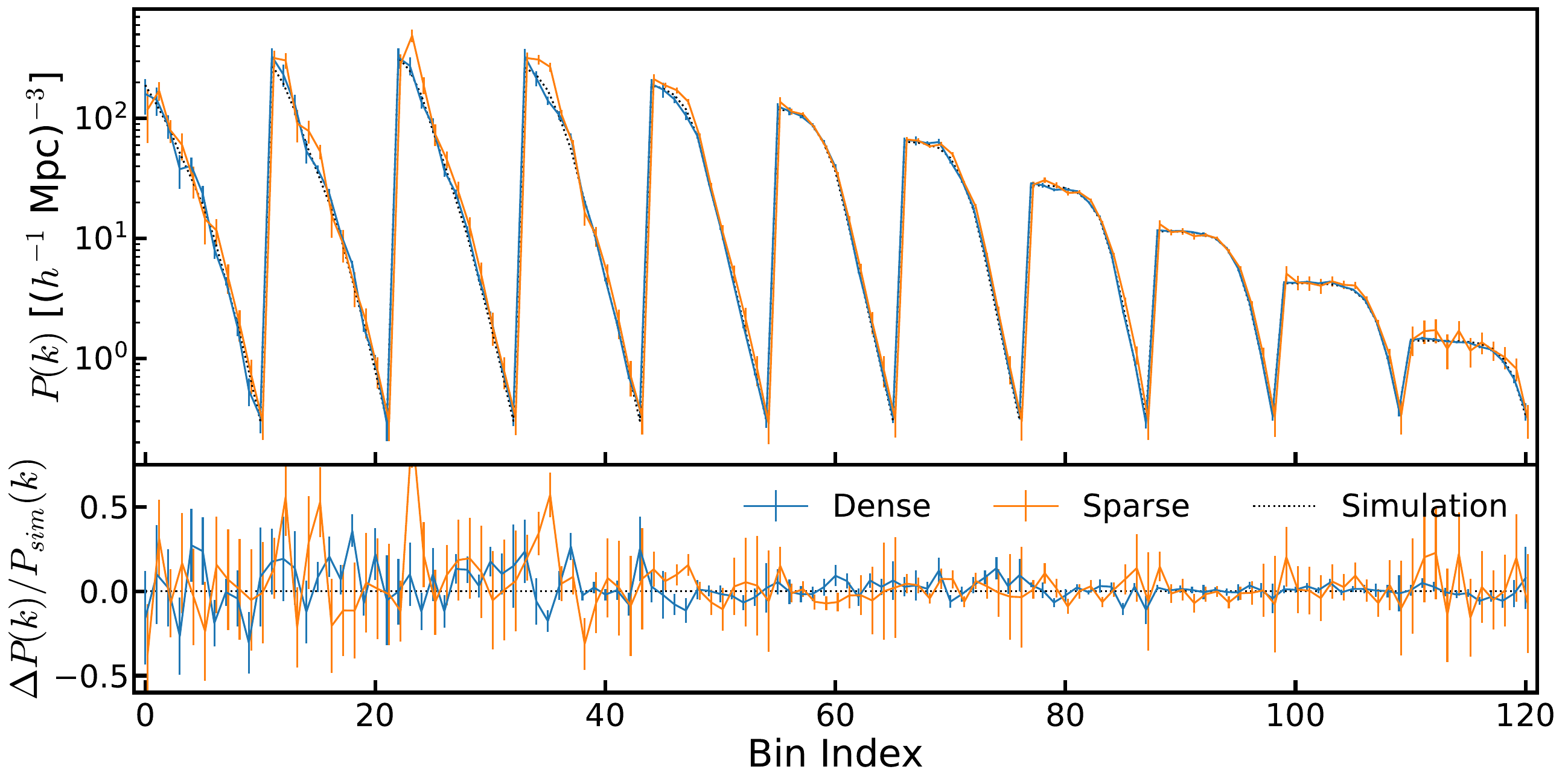}
    \vspace{-0.1in}
    \caption{Reconstructed band-powers and associated errors from the dense and sparse mock catalogs shown as a function of bin index. Bins are ordered by $k_\perp$ first, and then by $k_\parallel$. The bottom panel shows the relative residuals with respect to the theory input power spectrum. We find good agreement across a range of scales with reasonable error estimates.}
    \label{fig:p3d}
    \vspace{-0.1in}
\end{figure*}

We use two survey configurations: (i) sparse which is inspired by the DESI number density of $\sim 30$ sources per $\mathrm{deg}^2$ which roughly follows the noise/survey properties of \citet{2020MNRAS.497.4742K}. We assume constant resolution power of $R = 3200$ and a constant signal-to-noise per angstrom of 2 (corresponding to $\sigma_F=0.7$ per resolution element). (ii) The dense configuration is inspired by the Prime Focus Spectrograph Galaxy Evolution Survey \citep{2022PFSGE}, which targets both QSOs and Lyman Break Galaxies at greater depth ($g<24.2$) although at a comparable spectral resolution as DESI. This broader targeting will allow a significantly greater number density of $\sim$1000 background sources per deg$^2$. For this larger and denser sample we assume a signal-to-noise distribution similar to that found in the CLAMATO survey \citep{2018CLAMATO1,2022CLAMATODR2}. We assume the S/N is drawn from a power-law distribution with minimum value S/N$_\textrm{min}=1.0$ (i.e.\ $\rm{d}n_\textrm{ los}/\rm{d}(\textrm{S/N}) \propto \textrm{S/N}^{\alpha}$) with spectral amplitude $\alpha=2.7$. We also assume a maximal amplitude $\textrm{S/N}_\textrm{max}=10.0$ due to detector saturation on the bright end. We emphasize that we are not providing forecasts for both surveys but emulate \lya-specific survey geometries and number densities. 

We choose a highly anisotropic angular selection function to mimic irregularities in survey structure, including bright star masks, un-observed stripes, and deep fields. We pad the edges by $10\hinvMpc$ to avoid edge-effects due to the periodic nature of the Fourier operations used in our reconstruction. Note that we choose the same geometry for the dense and sparse fields, shown in Fig.~\ref{fig:angular_selection}. Additionally, we assume that background sources are uniformly distributed radially in the volume, resulting in a linear redshift distribution of data pixels. To increase the level of realism, we include Damped Lyman Alpha systems (DLAs) where the flux is damped in a region of 15 angstroms (with no explicit modelling of their damping wings or other features).\footnote{Note that we make the simplifying assumption that DLAs are uncorrelated with the underlying structure.} Each sight-line has a 10\% chance of having a DLA. For each DLA region, we mask out the DLA in noise, i.e. we increase the noise in those regions to very large values. This effectively removes them from the likelihood analysis.

\section{Results}
\label{sec:results}

\begin{figure*}
    \centering
    \includegraphics[width=1\linewidth]{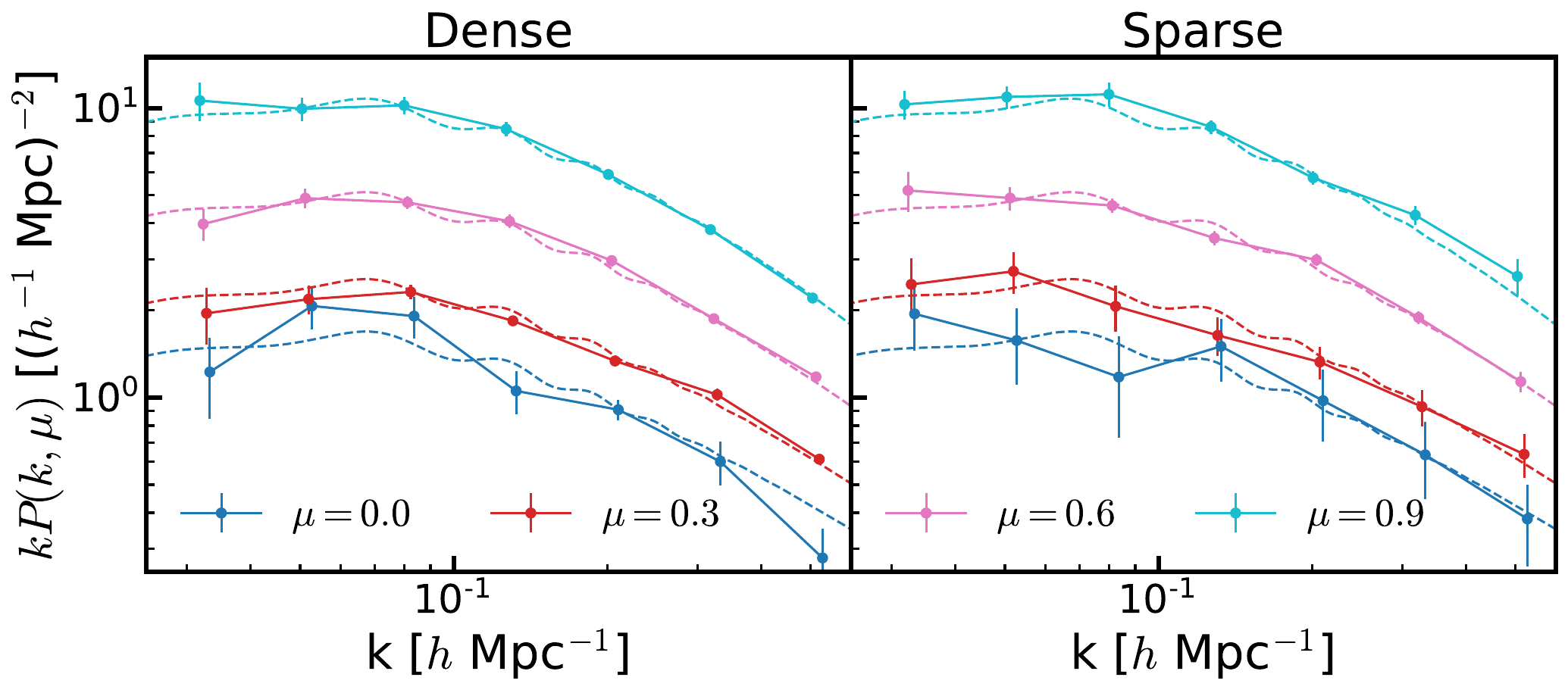}
    \vspace{-0.1in}
    \caption{Projection of the reconstructed power spectrum and covariance from $\{k_\perp,k_\parallel\}$ space to $\{k,\mu\}$ space. The error bars are taken from the diagonal of the square root of the covariance matrix in $\{k_\perp,k_\parallel\}$ space and also projected. We perform this projection by interpolating in $\{k_\perp,k_\parallel\}$ space and then outputting values along lines of constant $\mu$. The left (right) panel shows results from the dense (sparse) mock catalog. There is good agreement between the underlying model and our reconstruction.}
    \label{fig:p_proj}
    \vspace{-0.1in}
\end{figure*}

In this section, we present the results of our band-power and covariance estimator on GRFs with a fiducial Kaiser power spectrum with a non-linear correction term obtained from hydrodynamical simulations \citep{2003ApJ...585...34M}. We use the approach described in Section \ref{sec:methods}, starting from band-powers normally distributed around their true values, with a standard deviation of $0.3$ times the fiducial power spectra, $P_{\rm F}(k,\mu)$. For each optimization step, we average over 40 simulations with a convergence criteria of $\theta_{\rm tol} = 10^{-5}$. This takes about 2 hours (approximately $2000$ steps) on a 1 GPU (NVIDIA A100-SXM 40Gb) core including the Hessian estimation.

We show the band-power results of our reconstruction in Fig.~\ref{fig:p3d} with  errors from the reconstructed covariance matrix using Eq.~\ref{eq:H}. To generate an intuitive understanding for the reader, we also show the projection into $\{k,\mu\}$ space in Fig.~\ref{fig:p_proj} for both configurations. In both, reconstructed space and projected space, we find reasonable error estimates which generally follow a Gaussian distribution. In our reconstructed space, we find a reduced chi-squared, $\chi^2$, of 1.31 and 1.11 for the dense and sparse configurations, respectively. 

\begin{figure}
    \centering
    \includegraphics[width=1.00\linewidth]{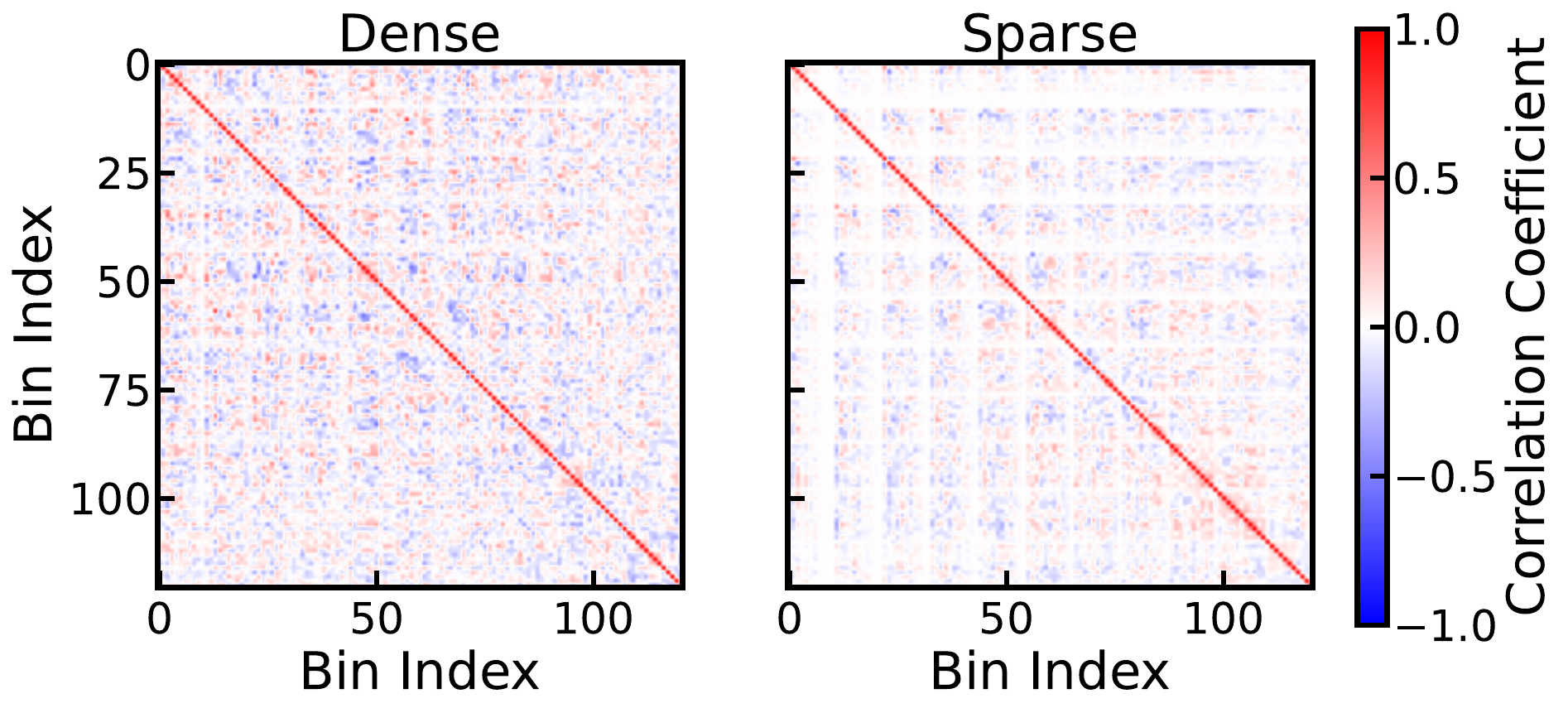}
    \caption{Correlation matrix of the reconstructed band-powers from the dense and sparse catalogs for the particular geometry shown in Fig.~\ref{fig:angular_selection}. Lack of off-diagonal correlations present in certain band-powers in the sparse catalog indicates lack of constraining power in that bin due to survey properties, i.e. the prior dominates the resulting likelihood. The grid structure is to be expected as the bins are ordered by $k_\perp$ and then by $k_\parallel$ (see Figure \ref{fig:p3d}).}
    \label{fig:cm}
\end{figure}

The corresponding correlation matrix is shown in Fig.~\ref{fig:cm}. This matrix is averaged over 20 realizations to reduce numerical noise. The correlation matrix is defined as
\begin{equation}
\label{eq:corr_mat}
    \mathbf{R}^{ij} \equiv \frac{\Sigma_{ij}}{\sqrt{\Sigma_{ii}\Sigma_{jj}}} \, ,
\end{equation}
and by construction it is unity along the diagonal. The covariance matrix is denoted by $\Sigma$ of power spectra in bins $i, j$. We find relatively small off-diagonal contributions which arise from the complex survey geometry. As expected, these contributions have higher amplitude in the dense survey as the data is more informative. The particularities of the survey geometry impact the specific $\mu$ dependencies in the error-bars and may not be generic for all surveys.

\section{Conclusions \& Outlook} \label{sec:conclusion}

In this work, we have presented \texttt{MAPLE}, a Wiener-filter-based optimization \lya three-dimensional power spectrum and full covariance/window function for two realistic survey configurations. This method is fast and scalable, and we make it publicly available on GitHub. With this method we find a consistent reconstructed result with good reduced $\chi^2$ error properties.

In this work, we choose to optimize each individual band-power on a rectangular grid, as it is the basis of the transfer function. This choice allows greatest flexibility and independence, as it can be used to constrain any P3D model. However, depending on the analysis of interest this might not be optimal. Also, one can directly use any differentiable parameterized model of $k_\perp$ - $k_\parallel$ to directly infer model parameters from \lya Forest data. This could include directly using Eq.~\eqref{eq:p3d} (if given a differentiable model for $P_{\rm L}(k)$ with respect to cosmological parameters), using an emulator for the whole P3D distribution, analytical P3D models \citep{2024PhRvD.109b3507I}, interpolation over multipole moments, etc.

The main goal of this letter is to demonstrate a promising new approach to measuring the P3D for \lya forest data. While our analysis choices are idealized and need to be tested on realistic mocks prior to application on \lya data, we have shown that the method is fast, scalable, and can be used to estimate the P3D and its covariance. While beyond the scope of this work, the underlying framework is flexible and allows joint modelling/marginalization over survey systematics. For example, one can jointly optimize the band-powers with QSO continuum estimation (e.g. PCA components, i.e. \citet{2012AJ....143...51L}) or DLA identification allowing for consistent propagation of uncertainties throughout the data analysis pipeline-line into the final cosmological analysis. This is particularly important for modes along the line of sight, which have significant correlated error contributions. 

An alternative approach to cosmological inference from \lya forest data would be a dynamical reconstruction approach where the initial matter density phases are reconstructed as opposed to the late time flux field  \citep{2019TARDISI,2021TARDISII}. While these approaches have been used successfully for large-scale structure inference \citep{2022CLAMATODR2}, applying the techniques in this paper to cosmological inference would be difficult due to the complex hydrodynamical modelling. In particular, uncertainty in the hydrodynamics must be accounted for in a differentiable way in order for the underlying optimization to be computationally tractable. This could be done via parameterized models \citep{2022constrainfgpa} or latent space deep learning surrogate modelling \citep{2022hyphy}. We leave this full analysis to future work.

\section*{Acknowledgements}
We thank Pat McDonald, Andreu Font-Ribera, Julien Guy, Nathalie Palanque-Delabrouille, Uro\v{s} Seljak, and Marius Millea for their helpful insights. 

This work was partially supported by the DOE's Office of Advanced Scientific Computing Research and Office of High Energy Physics through the Scientific Discovery through Advanced Computing (SciDAC) program.
This research used resources of the National Energy Research Scientific Computing Center, a DOE Office of Science User Facility supported by the Office of Science of the U.S. Department of Energy under Contract No. DEC02-05CH11231.

%%%%%%%%%%%%%%%%%%%%%%%%%%%%%%%%%%%%%%%%%%%%%%%%%%
\section*{Data Availability}
The data underlying this article will be shared on reasonable request. We make our code implementation publicly available\footnote{\url{https://github.com/bhorowitz/MAPLE}}.



\bibliographystyle{mnras}
\bibliography{example,mypapers} % if your bibtex file is called example.bib

%\begin{thebibliography}{99}
%\end{thebibliography}

%%%%%%%%%%%%%%%%%%%%%%%%%%%%%%%%%%%%%%%%%%%%%%%%%%

%%%%%%%%%%%%%%%%% APPENDICES %%%%%%%%%%%%%%%%%%%%%

%\appendix



%%%%%%%%%%%%%%%%%%%%%%%%%%%%%%%%%%%%%%%%%%%%%%%%%%

% Don't change these lines
\bsp	% typesetting comment
\label{lastpage}
\end{document}